# Lorentz transformations with arbitrary line of motion

**Chandru Iyer[1] and G. M. Prabhu[2]**

[1]Techink Industries, C-42, phase-II, Noida, India
[2]Department of Computer Science, Iowa State University, Ames, IA, USA
Contact E-mail: prabhu@cs.iastate.edu

**Abstract**
Sometimes it becomes a matter of natural choice for an observer (A) that he prefers a coordinate system of two-dimensional spatial x-y coordinates from which he observes another observer (B) who is moving at a uniform speed along a line of motion, which is not collinear with A's chosen x or y axis. It becomes necessary in such cases to develop Lorentz transformations where the line of motion is not aligned with either the x or the y-axis. In this paper we develop these transformations and show that under such transformations, two orthogonal systems (in their respective frames) appear non-orthogonal to each other. We also illustrate the usefulness of the transformation by applying it to three problems including the rod-slot problem. The derivation has been done before using vector algebra. Such derivations assume that the axes of K and K' are parallel. Our method uses matrix algebra and shows that the axes of K and K' do not remain parallel, and in fact K and K' which are properly orthogonal are observed to be non-orthogonal by K' and K respectively.

## 1. Introduction

In most treatments on special relativity [1, 2, 3], the line of motion is aligned with the x-axis. This is a natural choice because in such a situation the (y, z) coordinates are invariant under the Lorentz transformations (y' = y and z' = z). However, it is of interest to study the case when the line of motion does not coincide with any of the coordinate axes. Practical instances of such a situation are an airplane during landing or take off. The ground at the airfield has a natural coordinate system with the x-axis parallel to the ground, whereas the airplane ascends or descends at an angle with the ground. In this paper we develop such a formulation for a two-dimensional spatial system which can also be extended to three-dimensional spatial systems.

The pedagogical contribution of this paper lies in the derivation of the Lorentz transformation when the line of motion is not aligned with any of the axes. In 2-dimensional space the algebra is manageable as demonstrated by its application to the rod-slot problem. Since the algebra becomes more complex in 3-dimensional space, the student will realize the advantages of aligning one of the coordinate axes with the line of motion and appreciate why this is the accepted practice.

The results of this article have been derived earlier by using vector algebra [6]. However, such derivations assume that the axes of K and K' are parallel. Our method uses matrix algebra and shows that the axes of K and K' do not remain parallel, and in fact K and K' which are properly orthogonal are observed to be non-orthogonal by K' and K respectively. Since the result is not in conformity with the assumption (by vector algebra method) of parallelity of the axes of K and K' and the matrix method adopted by us does not make any such assumption, we reckon that the matrix method is more consistent and also more easily usable as illustrated by application to three problems.

## 2. Development of generalized 2-d Lorentz transformations

The transformation matrix for planar rotation by an angle θ in the anti-clockwise direction is given by

$$\begin{pmatrix} \cos\theta & \sin\theta \\ -\sin\theta & \cos\theta \end{pmatrix}$$

So  x' =  x cos θ  + y sin θ
    y' =  − x sin θ  + y cos θ

Within an inertial frame, when we add the time coordinate and consider the fact that a spatial rotation has no effect on the time coordinate, we add the equation t' = t, thus getting the transformation for spatial rotation to be:

$$\begin{pmatrix} x' \\ y' \\ t' \end{pmatrix} = \begin{pmatrix} \cos\theta & \sin\theta & 0 \\ -\sin\theta & \cos\theta & 0 \\ 0 & 0 & 1 \end{pmatrix} \begin{pmatrix} x \\ y \\ t \end{pmatrix} \quad (1)$$

The last row of the matrix indicates t' = t, and the two zeros in the last column of the matrix indicates there is no cross effect of t on x' or y'.

We use the symbol $R_\theta$ to denote this transformation, indicating spatial rotation anticlockwise by an angle θ.

Similarly, the Lorentz transformation along the x-axis yields the following matrix in a two-dimensional spatial system:

$$\begin{pmatrix} x' \\ y' \\ t' \end{pmatrix} = \begin{pmatrix} \gamma & 0 & -v\gamma \\ 0 & 1 & 0 \\ -v\gamma/c^2 & 0 & \gamma \end{pmatrix} \begin{pmatrix} x \\ y \\ t \end{pmatrix} \quad (2)$$

where $\gamma = \dfrac{1}{\sqrt{1 - v^2/c^2}}$.

The middle row of the matrix, (0, 1 , 0), indicates y' = y, and the two zeros in the middle column of the matrix indicate that there is no cross effect of y on x' or t'. We use the symbol $L_{xv}$ to denote a Lorentz transformation of magnitude v along the x-axis.

When the line of motion is inclined at an angle θ with the x-axis, we can use the transformation $R_{(-\theta)} L_{xv} R_\theta$ between two frames K and K' which observe each other moving at a velocity of +v and −v respectively as shown in Figure 1. As observed by both the frames, the proper angle of the line of motion is θ with respect to their respective x-axes. Noting that cos (−θ) = cos θ and sin (−θ) = − sin θ, we obtain the matrix A for $R_{(-\theta)} L_{xv} R_\theta$:

$$A = \begin{pmatrix} \gamma \cos^2\theta + \sin^2\theta & \sin\theta.\cos\theta(\gamma-1) & -v\gamma\cos\theta \\ \sin\theta.\cos\theta(\gamma-1) & \gamma\sin^2\theta + \cos^2\theta & -v\gamma\sin\theta \\ \dfrac{-v\gamma\cos\theta}{c^2} & \dfrac{-v\gamma\sin\theta}{c^2} & \gamma \end{pmatrix} \quad (3)$$

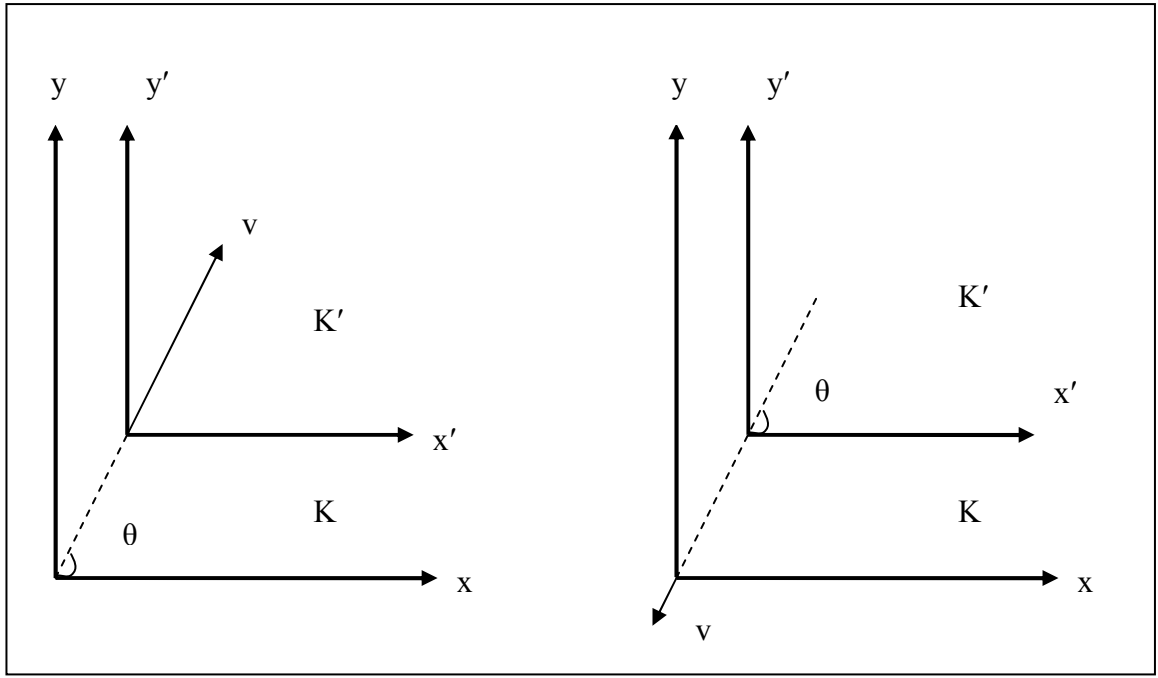

**Figure 1.** Relative motion of K (on left) and K' (on right)

The above matrix A transforms the event coordinates (x, y, t) of frame K to (x', y', t') of frame K' when K and K' are moving at a proper angle θ (anti clockwise with respect to each other's positive x, x' –axis respectively) and the relative velocity of K' with respect to K is +v.

The inverse of matrix A, denoted as B, is obtained by substituting –v for v:

$$B = \begin{pmatrix} \gamma \cos^2\theta + \sin^2\theta & \sin\theta.\cos\theta(\gamma-1) & v\gamma\cos\theta \\ \sin\theta.\cos\theta(\gamma-1) & \gamma\sin^2\theta + \cos^2\theta & v\gamma\sin\theta \\ \dfrac{v\gamma\cos\theta}{c^2} & \dfrac{v\gamma\sin\theta}{c^2} & \gamma \end{pmatrix} \quad (4)$$

The matrix B transforms the event coordinates (x', y', t') of frame K' to (x, y, t) of frame K, when K and K' are moving at a proper angle θ (anti clockwise with respect to each other's positive x, x' –axis respectively) and the relative velocity of K with respect to K' is –v. By using standard trigonometric identities, it can be verified that AB = BA = I, the identity matrix.

To visualize the x'-axis as seen by K we let y' = 0 and t = 0 (this is an instant in K, all points with y' = 0). For this condition to hold we obtain from K' = AK the following equation:

$a_{21} * x + a_{22} * y = 0$ (5)

The slope of this line is $- a_{21}/a_{22} = \sin\theta.\cos\theta (1 – \gamma) / [\gamma \sin^2\theta + \cos^2\theta]$.

One can verify that this line has a negative slope and is not aligned with the x-axis; this is notwithstanding the fact that the proper angle between the line of motion and the x-axis is the same in K and K'. Further we can infer (from the fact that this slope is not dependent on the sign of v) that the slope of the x'-axis as observed from K is the same as the slope of the x-axis as observed from K'.

Similarly the slope of the y' and y axes as observed from K and K' are also negative. Figure 2 depicts the manner in which two orthogonal systems appear non-orthogonal to each other. In general, the quadrants of the 'moving' frame through which the line of motion (drawn through the origin) passes, expand; the other two quadrants of the 'moving' frame contract, as observed by an observer co-moving with the 'stationary' frame.

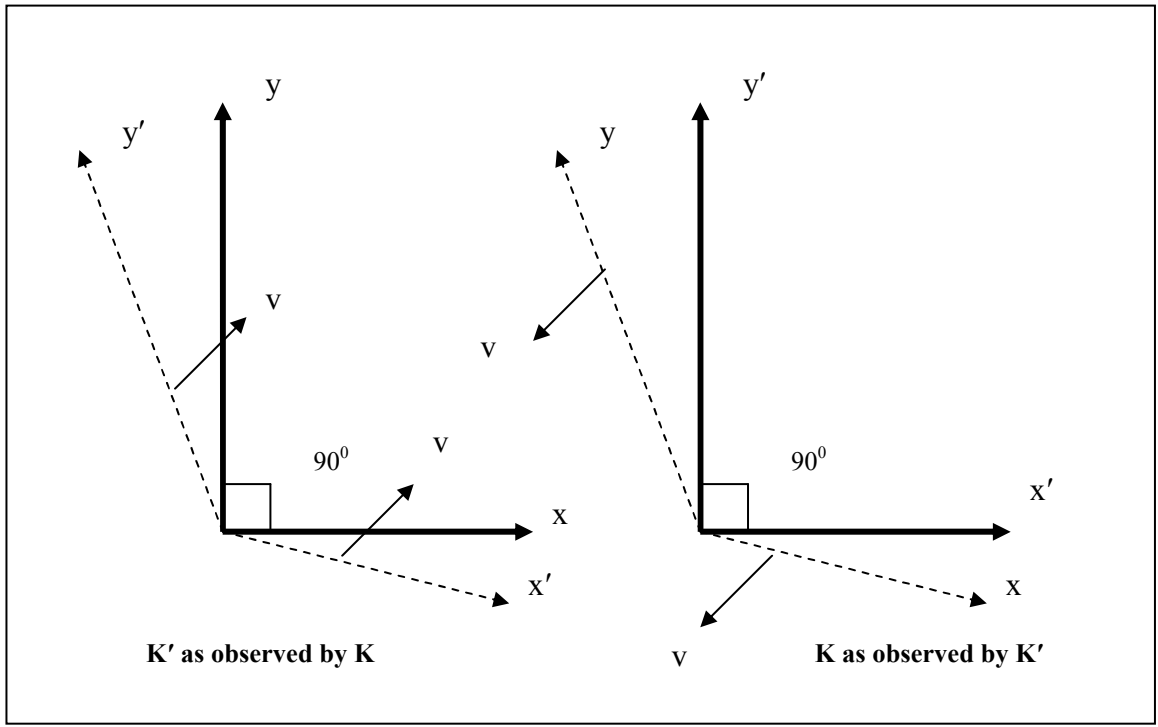

**Figure 2.** Observations from K (on left) and K' (on right):
Orthogonal proper systems appear non-orthogonal

## 3. Application of the transformation to an object moving at an incline with respect to a rod

Consider a rod of proper length L. Let us say an object is moving at a velocity v and at an angle θ with the axis of the rod (as observed from the co-moving inertial frame of the rod). What will be the apparent length of the rod as observed by an observer co-moving with the inertial frame of the object?

We have a preferred coordinate system (F) co-moving with the rod as shown in Figure 3. In this system the x axis is along the length of the rod and one end of the rod has coordinates x = 0; y = 0. The other end of the rod has coordinates x = L; y = 0.

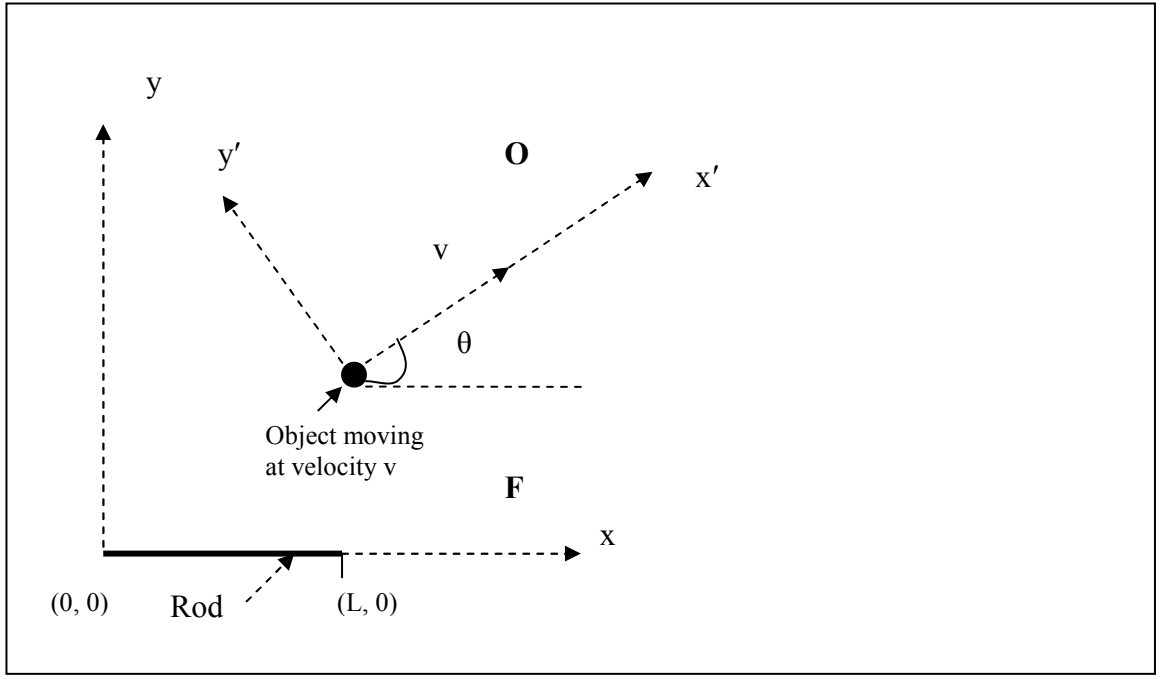

**Figure 3.** Rod in Frame F and object in frame O moving at velocity v

The space-time coordinates of the two end points of the rod at any arbitrary time t read (0, 0, t) and (L, 0, t). We can also visualize a coordinate system (O) co-moving with the object with the x' axis aligned with the line of relative motion.

In order to transform from F to O we use the transformation $[L_{xv} R_\theta]$ as specified in equations (1) and (2). The resultant transformation is

$$T = L_{xv} R_\theta = \begin{pmatrix} \gamma & 0 & -v\gamma \\ 0 & 1 & 0 \\ -v\gamma/c^2 & 0 & \gamma \end{pmatrix} \begin{pmatrix} \cos\theta & \sin\theta & 0 \\ -\sin\theta & \cos\theta & 0 \\ 0 & 0 & 1 \end{pmatrix}$$

$$= \begin{pmatrix} \gamma \cos\theta & \gamma \sin\theta & -v\gamma \\ -\sin\theta & \cos\theta & 0 \\ \dfrac{-v\gamma \cos\theta}{c^2} & \dfrac{-v\gamma \sin\theta}{c^2} & \gamma \end{pmatrix} \qquad (6)$$

We can apply matrix T as derived in equation (6) to transform the event coordinates from F to O. One end of the rod at t = 0 in F has the coordinates (0, 0, 0) and it transforms to (0, 0, 0) in O. This corresponds to t' = 0. In order to observe the other end of the rod from the inertial frame O at t' = 0, we transform the coordinates of the other end of the rod at some instant t in frame F and set t' = 0.

$$\begin{pmatrix} x' \\ y' \\ 0 \end{pmatrix} = T \begin{pmatrix} L \\ 0 \\ t \end{pmatrix} \qquad (7)$$

From equation (7) we obtain

x' = L γ cos θ – v γ t  (8)
y' = – L sin θ  (9)
0 = – (Lv γ cos θ )/$c^2$ + γt  (10)

Solving equation (10) for t and substituting its value in equation (8) gives x' = L cos θ / γ .

Thus at the instant t' = 0 in O, one end of the rod has coordinates x' = 0; y' = 0, and the other end has coordinates x' = L cos θ / γ and y' = – L sin θ. Therefore the apparent length of the rod as observed by O is $L* \sqrt{(\cos^2\theta / \gamma^2) + \sin^2\theta}$. This quantity is equal to L when γ = 1 (no relative motion), is equal to L/γ when θ = 0 (traditional Lorentz transformation along line of motion) and is equal to L when θ = 90 (no variation in length perpendicular to line of motion). All these results are consistent with known concepts of Lorentzian contraction.

**4. Application of the transformation to the collision of inclined rods problem**

Another suitable problem for application is the "collision of the inclined rods" as described in [4]. Here we have two parallel (considering proper angles) rods in relative motion with the axis of both the rods making a proper angle θ with the line of relative motion. If both the rods select their respective axis as their x and x' axis we can use the transformations A and B in equations (3) and (4) as the forward and inverse transformations between the two inertial frames.

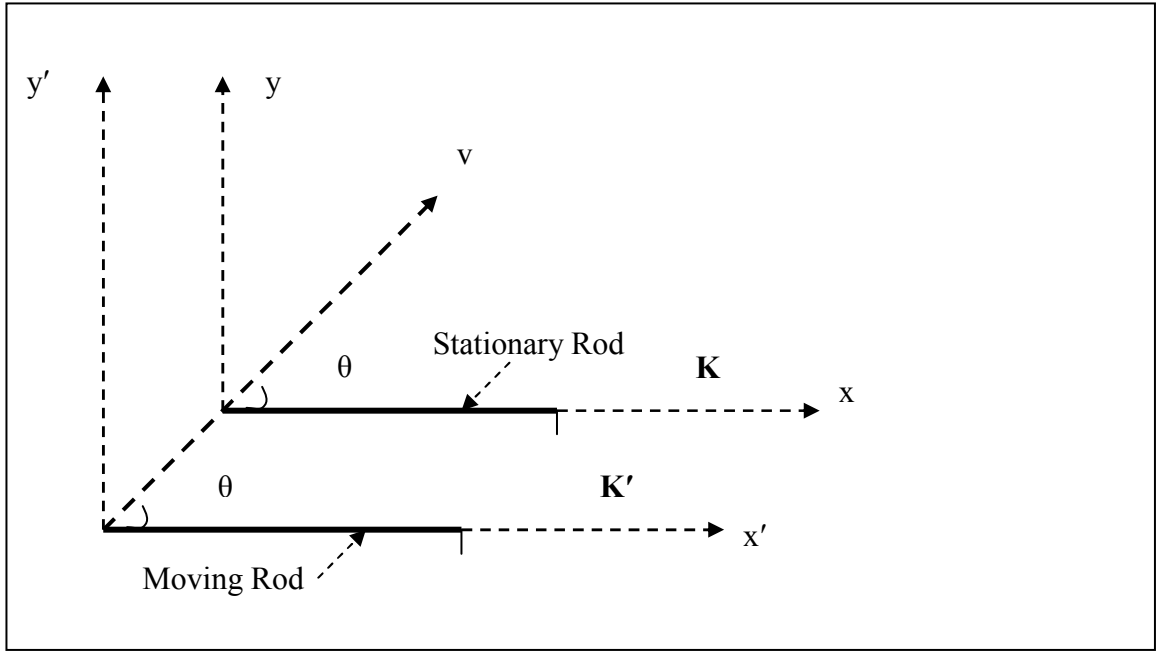

**Figure 4.** Rod in Frame K′ moves towards stationary rod in Frame K at velocity v

In Figure 4, the bottom collision is taken as the space-time origin (0, 0, 0) in both the frames K and K′ [4]. The top collision has the coordinates (L, 0, t) in frame K and (L, 0, t') in frame K'. Using matrix A from equation (3) to transform (L, 0, t) and equating the result to (L, 0 ,t') we obtain the following equations.

$$(\gamma \cos^2 \theta + \sin^2 \theta) L - (v \gamma \cos \theta) t = L \tag{11}$$

$$L \sin \theta \cos \theta (\gamma - 1) - (v \gamma \sin \theta) t = 0 \tag{12}$$

$$-(v\gamma/c^2) L \cos \theta + \gamma t = t' \tag{13}$$

From equations (11) and (12) we get the same result, namely $t = (L/v) \cos \theta [1 - (1/\gamma)]$. This redundancy is due to our prior assumption that the top end of the first rod collides with the top end of second rod. Substituting this into equation (13) yields $t' = (L/v) \cos \theta [(1/\gamma) - 1]$. This is consistent with the results in [4] and confirms the reversal in time order of the top and bottom collisions in the two inertial frames co-moving with each of the two rods respectively.

## 5. Application of the transformation to the rod and slot problem

Another suitable application for the transformation derived above is the rod and slot problem described in [5]. Here the proper angles are unequal and this presents a more general application. In this scenario the rod exhibits motion in two directions, but only with constant velocity. Furthermore, the line of motion (that is, the line joining the centers of the rod and slot) is not aligned with either the axis of the rod or the slot, there is no gravity, and thus no stress or propagation of stress.

We assume the proper angle of the axis of the rod with the line of motion to be Φ and that of the slot to be α. The rod has a coordinate system with the x-axis along the axis of the rod and the origin at the center of the rod. Similarly the slot has a coordinate system with the x-axis along the axis of the slot and the origin at the center of the slot. The meeting of the centers of the rod and slot is the space-time origin with (x, y, t) = (x', y', t') = (0, 0, 0) as shown in Figure 5. The rod is shown parallel to the slot in Figure 5 ($\Phi = \alpha$), but in general $\Phi$ does not have to be equal to $\alpha$.

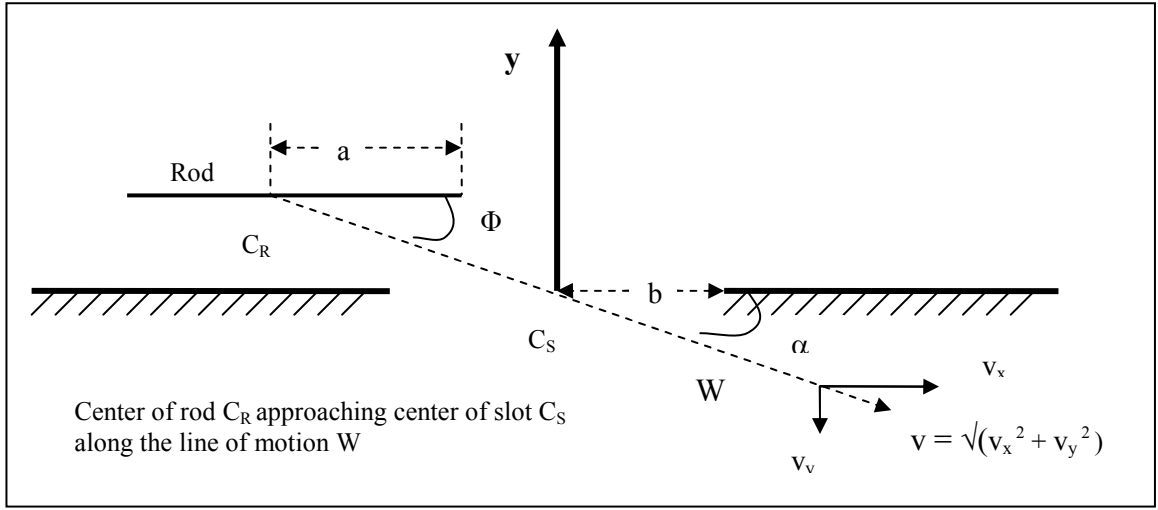

**Figure 5.** Initial conditions of the rod-slot problem

When we transform from the rod's reference frame to the slot's reference frame, we have to use the following transformations in sequence. $R_{(-\Phi)}$ will be the first transformation to align the line of motion with the x axis. Then we do a Lorentz transformation of magnitude $-v$ to switch to the slot's reference frame; this is designated as $L_{x(-v)}$. The resultant coordinates are in the slot's reference frame but with the x-axis aligned with the line of motion. Since the slot has a preferred system with the x-axis aligned with the slot's axis, we need to perform a transformation $R_{(\alpha)}$. Thus the overall transformation is $R_{(\alpha)} L_{x(-v)} R_{(-\Phi)}$. Denoting this transformation as D, we obtain the elements of D as follows:

$$D = \begin{pmatrix} \gamma \cos\phi \cos\alpha + \sin\alpha \sin\phi & -\gamma \sin\phi \cos\alpha + \sin\alpha \cos\phi & v\gamma \cos\alpha \\ -\gamma \cos\phi \sin\alpha + \cos\alpha \sin\phi & \gamma \sin\phi \sin\alpha + \cos\phi \cos\alpha & -v\gamma \sin\alpha \\ \dfrac{v\gamma \cos\phi}{c^2} & \dfrac{-v\gamma \sin\phi}{c^2} & \gamma \end{pmatrix} \quad (14)$$

Let $a$ denote half the length of the rod, and $b$ denote half the length of the slot. The leading edge of the rod has the coordinates $x = a$; $y = 0$ for any arbitrary time t in the rod's frame. This can be denoted by the triple $(a, 0, t)$. Similarly the front edge of the slot has the coordinate $(b, 0, t')$ in the slot's frame for any arbitrary time t' in the slot's frame. In order for the rod to just pass through the slot, $(a, 0, t)$ must transform to $(b, 0, t')$ for some value of t.

Thus $\begin{pmatrix} b \\ 0 \\ t' \end{pmatrix} = D \begin{pmatrix} a \\ 0 \\ t \end{pmatrix}$

This gives rise to the following three equations:

$$a \gamma \cos\Phi \cos\alpha + a \sin\alpha \sin\Phi + vt \gamma \cos\alpha = b \quad (15)$$

$$-\, a\, \gamma\, \cos\Phi \sin\alpha + a\, \cos\alpha\, \sin\Phi -\ vt\, \gamma\, \sin\alpha = 0 \qquad (16)$$

$$(a\, v\, \gamma\, /\, c^2)\, \cos\Phi + \gamma\, t\ =\ t' \qquad (17)$$

From equation (16) we get
$t = (a\, \cos\alpha\, \sin\Phi - a\, \gamma\, \cos\Phi \sin\alpha) / (v\, \gamma\, \sin\alpha)$

Substituting for t in equation (15) we get

$a\, \gamma\, \cos\Phi \cos\alpha + a\, \sin\alpha\, \sin\Phi + (a\, \cos\alpha\, \sin\Phi - a\, \gamma\, \cos\Phi \sin\alpha) \cos\alpha / \sin\alpha = b$

Multiplying both sides by $\sin\alpha$ and simplifying, we obtain

$a\, \sin\Phi\, (\sin^2\alpha + \cos^2\alpha) = b\, \sin\alpha$

or   $a\, \sin\Phi = b\, \sin\alpha$

Thus we find that in order for the rod to just pass through the slot, we obtain a condition that depends only on the four proper quantities, that is, the proper lengths of the rod and slot and the proper angles between the line of motion and the axis of the rod and slot. Further, it is easy to deduce that whenever $a\, \sin\Phi < b\, \sin\alpha$, the rod passes through the slot.
  It can be further deduced from the above discussions that if the line of motion passing through the center of the rod does not intersect the slot at its center, but divides the slot into two unequal lengths $b_1$ and $b_2$, then the condition for the rod to pass through the slot is given by

($a\, \sin\Phi < b_1\, \sin\alpha$) AND ($a\, \sin\Phi < b_2\, \sin\alpha$).

## 6. Summary

We derived a general Lorentz transformation in two-dimensional space with an arbitrary line of motion. We applied it to three problems and demonstrated that it leads to the same solution as already established in the literature. The solved problems using the transformation equations illustrate the convenience of the chosen coordinate system in either frame of reference. Thus, in certain situations, we see the merit of using the Lorentz transformations with the line of motion not coinciding with any of the coordinate axes. Our method can also be conceptually extended to 3-dimensional space with an arbitrary line of motion. However, the algebra is a lot more complex in this case. This may be one reason that conventionally the Lorentz transformations are performed with the line of motion coinciding with one of the coordinate axes.

Appendix - Reply to Comment by Tjiang and Sutanto
          Chandru Iyer and G.M. Prabhu [Eur J. Phys. **28** (2007) L15-L16]
              http://www.iop.org/EJ/abstract/0143-0807/28/3/N03

The comment received from Dr. Tjiang and Dr. Sutanto [Eur J. Phys. **28** (2007) L11-L14] http://www.iop.org/EJ/abstract/0143-0807/28/3/N02 is appreciated. We thank them for the keen interest they have shown to the paper. In reply to the points raised we state as under.

(1) The authors of the comment have stated that we have applied equation (2), which is the equivalent of equation (3) of our paper, to the rod–slot problem. This is incorrect. We have applied equation (14) of our paper to the rod–slot problem. Equation (14) of our paper has 3 parameters. The authors have expanded equation (3) of our paper, which has 2 parameters, to its 3-dimensional version and the result has 3 parameters. If equation (14) of our paper is expanded to its 3-dimensional version, it will have 6 parameters.

(2) We have pointed out in Figure 2 of our paper that when the two frames are in relative motion with the line of motion not coinciding with either axis, the axes of the frames cannot become parallel. The moving frame's axes are observed to be non-orthogonal. In such a situation, realizing the condition of having the axes of S and S′ being parallel is a requirement whose fulfillment appears to be questionable. What we actually achieve in equation (3) of our paper is that the components of the unit vector along the line of motion take the same form in both S and S′.

(3) It is well established in the literature that vector addition in relativity is non-commutative [1, 2]. For example, if we add $u_i + v_j$ in that order and alternatively, $v_j + u_i$ in that order, under relativity we get the combined velocity to have a magnitude of $\sqrt{u^2 + v^2 - (u^2 * v^2 / c^2)}$ in both cases. But the angle that the line of velocity makes with the *x*-axis is different in the two cases [2].

(4) This was our primary motivation in suggesting an alternative approach to using vector algebra. Our pedagogy has been to construct the transformation by first rotating the *x-y* plane to align the *x*-axis with the line of motion, perform a conventional Lorentz transformation, and then rotate back the *x-y* plane as per requirement of the 'moving' inertial frame in relation to the orientation of the object(s) co-moving with it. We felt that this approach resulted in ease of application to some problems.

(5) The two-dimensional version has 3 parameters, Φ, α, and *v* given in equation (14) on Page 189 of our paper. In the most general 3-d solution, we have to rotate the *x-y* plane, then the *x-z* plane so as to align the *x*-axis with the line of motion. This involves two angular rotations. A conventional Lorentz is performed at this point and so far we have 3 parameters. We presume that at this point (in the general case), the inertial frame has reached the required state of relative motion but the orientation of the axis is not as per its needs in relation to the orientation of the objects co-moving with it. Therefore, we have to rotate the co-ordinate planes as per the requirements of the 'moving' frame. This operation in 3-d space in general requires 1 rotation each of the 3 co-ordinate planes. So we end up with 6 parameters in the 3-d version in our approach, which renders the algebra more complex than the 2-d version.